\documentclass[aps, prl, twocolumn, nofootinbib, superscriptaddress]{revtex4-2}
\usepackage{amsmath}
\usepackage{amssymb}
\usepackage{amsfonts}
\usepackage{booktabs, array}
\usepackage[T1]{fontenc}

\usepackage[caption=false]{subfig}
\usepackage{tikz}
\usepackage{graphicx} 
\usepackage{multirow}
\usepackage{makecell}

\definecolor{linkcolor}{rgb}{0.0,0.3,0.5}
\usepackage[
    hypertexnames=false,
    unicode,
    colorlinks=true,
    linkcolor=linkcolor,
    citecolor=linkcolor,
    filecolor=linkcolor,
    urlcolor=linkcolor,
    pdfusetitle
]{hyperref}
\usepackage{orcidlink}

\newcommand{\rmd}{\mathrm{d}}
\newcommand{\thorn}{\text{\th}}
\newcommand{\etal}{\textit{et al.\ }}

\newcommand{\Perimeter}{\affiliation{Perimeter Institute for Theoretical Physics, Waterloo, ON N2L2Y5, Canada}}

\graphicspath{./Plots}
\begin{document}
    \title{Quadratic Mode Couplings in Rotating Black Holes and Their
    Detectability}
    \author{Neev Khera \orcidlink{0000-0003-3515-2859}}
    \email{nkhera@tsinghua.edu.cn}
    \affiliation{Department of Physics, University of Guelph, Guelph, Ontario, Canada N1G 2W1}
    \affiliation{Department of Astronomy, Tsinghua University, Beijing 100084, China}
    \author{Sizheng Ma \orcidlink{0000-0002-4645-453X}}
    \email{sma2@perimeterinstitute.ca}
    \Perimeter
    \author{Huan Yang \orcidlink{0000-0002-9965-3030}}
    \email{hyangdoa@tsinghua.edu.cn} \affiliation{Department of Astronomy, Tsinghua University, Beijing 100084, China}
    \date{September 2024}
    \begin{abstract}
        Quadratic quasinormal modes encode fundamental properties of black hole spacetimes.
        They are also one of the key ingredients of nonlinearities of General
        Relativity in the ringdown stage of binary black hole coalescence. In this
        work, we classify all possible quadratic coupling channels of
        quasinormal modes for a generic Kerr black hole, and use a frequency-domain
        pseudospectral code with hyperboloidal slicing to calculate these couplings.
        After accounting for all the channels in systems with reflection
        symmetry, our results become consistent with those extracted from numerical
        simulations and time-domain fits. 
        This agreement provides a compelling example demonstrating the success of black hole second-order perturbation theory.
        We also explore potential applications of our calculations in future ringdown data analysis by carrying out a detectability survey for various quadratic modes. We find that a few of them are observationally relevant for third-generation
        ground-based detectors like Cosmic Explorer, as well as the space-borne detector
        LISA (Laser Interferometer Space Antenna). 
    \end{abstract}
    \maketitle


    \section{Introduction}
    \label{sec:introduction} Black hole (BH) ringdown after binary BH
    coalescence encodes key information about the spacetime of the final BH
    \cite{Berti:2009kk,Konoplya:2011qq,Berti:2018vdi}, as well as the binary's inspiral
    properties, such as the mass ratio, spins, etc~\cite{London:2014cma,London:2018gaq,Hughes:2019zmt,Cheung:2023vki,Pacilio:2024tdl,MaganaZertuche:2024ajz}.
    Theoretically, it is important to characterize various linear and non-linear
    signals in the ringdown stage and their connection to the inspiral
    parameters, as predicted by General Relativity (GR). In particular, understanding
    nonlinear signals not only enables direct tests of nonlinearities in GR, but
    also allows more theoretically meaningful ringdown fits at earlier times. This
    is particularly important for data analysis, as fitting the ringdown from
    earlier times generally leads to a higher signal-to-noise ratio (SNR).

    Among various BH nonlinearities
    \cite{Mitman:2022qdl,Cheung:2022rbm,Ma:2022wpv,Zhu:2024rej,Redondo-Yuste:2023seq,Cheung:2023vki,Zhu:2024dyl,May:2024rrg,Khera:2023oyf,Ma:2024qcv,Bucciotti:2024zyp,Bucciotti:2024jrv,Bourg:2024jme,Redondo-Yuste:2023ipg,Sberna:2021eui,BenAchour:2024skv,Perrone:2023jzq,Kehagias:2023ctr},
    quadratic quasinormal modes (QQNMs) stand out due to their conceptual simplicity
    and their potential observability using current ringdown fitting techniques \cite{Carullo:2019flw,Isi:2021iql,Siegel:2023lxl,Siegel:2024jqd,Capano:2021etf,Wang:2023xsy,Ma:2023vvr,Ma:2023cwe}.
    QQNMs have been identified by fitting numerical time-domain simulation data
    at infinity
    \cite{Mitman:2022qdl,Cheung:2022rbm,Ma:2022wpv,Zhu:2024rej,Redondo-Yuste:2023seq,Cheung:2023vki}
    and on the BH horizon \cite{Khera:2023oyf}. However, a complete and accurate
    characterization of the modes relies on frequency domain methods, which is the
    focus of this work.

    Along with an earlier work \cite{Ma:2024qcv} (hereafter Paper I), we develop
    two independent frequency-domain methods to handle QQNMs for {\it generic}
    Kerr BHs. In Paper I, we analytically reconstructed the metric and solved
    the frequency-domain second-order Campanelli–Lousto-Teukolsky equation
    \cite{Campanelli:1998jv} along a complex contour. Alternatively, in this
    paper, we adopt the reduced second-order Teukolsky equation \cite{Green:2019nam,soton469806}
    and use the pseudo-spectral method to numerically assemble its source term.
    Subsequently, we solve the equation spectrally in hyperboloidal slicing \cite{Zenginoglu:2011jz,Zenginoglu:2007jw,PanossoMacedo:2024nkw}.
    Both methods generate consistent results for the coupling coefficient of the
    prograde mode $\ell=m=2,n=0$ for BH spins $\chi \in [0,0.95]$. Furthermore, we extend our discussions in Paper I by providing a comprehensive classification of all possible mode-coupling
    channels that contribute to the same frequency of quadratic modes. This classification allows us to
    resolve the disagreement between our previous QQNM calculations in Paper I and numerical simulations. 


    Leveraging the recent ringdown surrogate model\footnote{See also
    \cite{MaganaZertuche:2024ajz}.} \cite{Pacilio:2024tdl}, we discuss potential applications of our calculations in future ringdown data analysis. As an example, we carry out a parameter survey to forecast the SNRs of several
    quadratic modes measured by Cosmic Explorer
    (CE) and Laser Interferometer Space Antenna (LISA) \cite{Baker:2019nia,Evans:2021gyd}, complementing the recent study by
    Yi \etal \cite{Yi:2024elj}. Our findings indicate that multiple QQNMs are relevant for future ringdown analysis. 
    
    Throughout this work we adopt the geometric unit by setting $G=c=1$. Complex
    conjugates are represented by overlines. We use $(l,m,n,p)$ to stand for a QNM, where $p={\rm sgn}({\rm Re}(\omega))$; and $\omega$ to denote frequency values.
    Due to the reflection symmetry of Kerr, QNMs come in pairs with frequency values related via
    \begin{align}
        \omega^{p}_{\ell mn}=-\bar{\omega}^{-p}_{\ell-mn}, \label{eq:qnm_kerr_symmetry}
    \end{align}
    where modes with $(m,p)$ and $(-m,-p)$ (and the same
    $l,n$) are mirrors of each other.

    \section{QQNMs with hyperboloidal slicing}
    QNMs are known to diverge at spatial infinity and at the bifurcate horizon.
    While this divergence poses no physical issue because they do not lie in the
    region where the QNMs are a good approximation, it can cause numerical challenges.
    In Paper I, we deal with the divergence by using the complex plane approach.
    Here, we adopt an alternative method---hyperboloidal time slicing---which
    avoids the problematic regions. This enables direct numerical treatments of
    the reduced second-order Teukolsky equation \cite{Green:2019nam,soton469806}
    \begin{align}
        \mathcal{T}\left[\Psi_{4}^{(2)}\right] = S_{4}^{(2)}. \label{eq:second_order_Teukolsky_eqn}
    \end{align}
    Using the hyperboloidal slicing, it has been shown that QNMs are eigenvectors
    of the non-normal time evolution operator~\cite{Jaramillo:2020tuu}. The
    separability of the Teukolsky equation can be used to find the eigenvectors
    in a more efficient way, as shown in~\cite{Ripley:2022ypi}.

    To obtain the source term $S_{4}^{(2)}$, we take a different approach from the
    analytic treatment used in Paper I---here, the source is computed
    numerically from the Einstein Tensor. Specifically, we consider the
    second-order perturbation of the metric
    \begin{equation}
        g_{ab}= g^{\circ}_{ab}+ \epsilon h_{ab}^{(1)}+ \epsilon^{2}h_{ab}^{(2)}+
        \mathcal{O}\left(\epsilon^{3}\right)\,,
    \end{equation}
    where $\epsilon$ is a book-keeping parameter that counts the orders. The Einstein
    Tensor, as a functional of the metric, can then be expanded as
    \begin{align}
        G_{ab}[g_{ab}] = & \epsilon G_{ab}^{(1)}[h_{ab}^{(1)}] + \epsilon^{2}G_{ab}^{(1)}[h_{ab}^{(2)}] + \epsilon^{2}G_{ab}^{(2)}[h_{ab}^{(1)}, h_{ab}^{(1)}] \notag \\
                         & +\mathcal{O}\left(\epsilon^{3}\right)\,,
    \end{align}
    where $G_{ab}^{(1)}$ and $2 G_{ab}^{(2)}$ are the first and second functional
    derivatives of the Einstein Tensor with respect to the metric. To first
    order, we take the GW strain to be a sum of QNMs. We solve for
    $h_{ab}^{(1)}$ by numerically computing the Hertz potential for QNMs in the
    Outgoing Radiation Gauge (ORG), using the separation technique described in~\cite{Ripley:2022ypi}.
    Note that for a QNM with frequency $\omega$, $h_{ab}^{(1)}$ will have both frequencies
    $\omega$ and $-\bar{\omega}$, which comes from the complex conjugations
    necessary to make $h_{ab}^{(1)}$ real.

    Next, by imposing Einstein's equations at second order, we get the equation
    \begin{equation}
        \label{eq:Einstein-2}G_{ab}^{(1)}[h_{ab}^{(2)}] = - G_{ab}^{(2)}[h_{ab}^{(1)}
        , h_{ab}^{(1)}]\,,
    \end{equation}
    thus $h_{ab}^{(2)}$ satisfies the linearized Einstein equations with
    $- G_{ab}^{(2)}[h_{ab}^{(1)}, h_{ab}^{(1)}]/8\pi$ acting as the effective
    stress-energy tensor at this order. After computing $G_{ab}^{(2)}[h_{ab}^{(1)}
    , h_{ab}^{(1)}]$ from the linearized metric $h_{ab}^{(1)}$, we convert it to
    the source $S_{4}^{(2)}$ for $\Psi_{4}^{(2)}[h_{ab}^{(2)}]$ in Eq.~\eqref{eq:second_order_Teukolsky_eqn}.
    Note that while Eq.~\eqref{eq:Einstein-2} only provides the source for the part
    of $\Psi_{4}$ that depends linearly on $h_{ab}^{(2)}$, in ORG quadratic
    contribution to $\Psi_{4}^{(2)}$ from $h_{ab}^{(1)}$ vanishes.

    Finally, we numerically solve the Teukolsky equations with no-incoming boundary
    conditions to obtain $\Psi_{4}^{(2)}$ and $\Psi_{0}^{(2)}$ for the quadratic
    modes. More details can be found in the Supplementary Material.

    \section{Classification of mode couplings}
    Quadratic coupling channels can be viewed in two perspectives. The first 
    traces the relationship from linear parent modes to quadratic daughter modes.
    Consider two linear QNMs $(\ell_{1},m_{1},n_{1},p_{1})$ and $(\ell_{2},m_{2},
    n_{2},p_{2})$, with strain amplitudes $A_{1}$ and $A_{2}$.
    The corresponding linearized metric $h_{ab}^{(1)}$ has four frequency components
    $\omega_{\ell_1m_1n_1}^{p_1}$, $-\bar{\omega}_{\ell_1m_1n_1}^{p_1}$, $\omega_{\ell_2m_2n_2}
    ^{p_2}$, and $-\bar{\omega}_{\ell_2m_2n_2}^{p_2}$, with the amplitudes $A_{1}$,
    $\bar{A}_{1}$, $A_{2}$ and $\bar{A}_{2}$, generated by the frequency pairs of the linear modes $(\ell_{1}, m_{1}, n_{1}, p_{1})$ and $(\ell_{2}, m_{2}, n_{2}, p_{2})$.
    Consequently, the source term in Eq.~\eqref{eq:Einstein-2} yields four quadratic
    frequencies:
    \begin{align}
        \omega_{\ell_1m_1n_1}^{p_1}+\omega_{\ell_2m_2n_2}^{p_2},        &  & \omega_{\ell_1m_1n_1}^{p_1}-\bar{\omega}_{\ell_2m_2n_2}^{p_2}, \notag        \\
        -\bar{\omega}_{\ell_1m_1n_1}^{p_1}+\omega_{\ell_2m_2n_2}^{p_2}, &  & -\bar{\omega}_{\ell_1m_1n_1}^{p_1}-\bar{\omega}_{\ell_2m_2n_2}^{p_2}, \notag
    \end{align}
    with source terms proportional to the products $A_{1}A_{2}$, $A_{1}\bar{A}_{2}$,
    $\bar{A}_{1}A_{2}$ and $\bar{A}_{1}\bar{A}_{2}$, respectively. This gives
    rise to {\it four} distinct channels through which two linear QNMs can
    generate QQNMs at different frequencies.

    The first perspective, which maps two parent modes to four daughter modes,
    is often practically inconvenient. Instead, it is more intuitive to adopt
    the opposite viewpoint: \emph{given a certain QQNM frequency, what are its parent
    modes?} To formalize this, we will use the following notation to denote a QQNM
    frequency component:
    \begin{align}
        \begin{pmatrix}\ell_{1},&m_{1},&n_{1},&p_{1}\\ \ell_{2},&m_{2},&n_{2},&p_{2}\end{pmatrix}. \label{eq:QQNM_notation}
    \end{align}
    Its time and azimuthal dependence is expressed as:
    \begin{align}
        \sim e^{-i(\omega_{\ell_1,m_1,n_1}^{p_1}+\omega_{\ell_2,m_2,n_2}^{p_2} ) t}e^{i(m_1+m_2)\phi}.
    \end{align}
    This notation represents the set of all QQNMs with that frequency, including various channels, and is not restricted to fixed daughter spherical/spheroidal harmonic.
    With this notation, several key observations can be made. First, QQNMs with
    this frequency only appear in the $(m_{1}+m_{2})$ GW harmonics due to
    the azimuthal angular selection rule. Second, different pairs of modes (up to exchange symmetry $1\leftrightarrow
    2$) generally produce distinct frequencies. An exception occurs for Schwarzschild
    BHs, where spherical symmetry makes QNM frequencies independent of $m$. Nonetheless,
    different $(m_{1},m_{2})$ pairs indicate distinct coupling channels. The total
    QQNM amplitude in a GW harmonic is the sum of all possible pairs $(m_{1},m_{2}
    )$ with $m_{1}+m_{2}$ fixed.

    The QQNM frequency in Eq.~\eqref{eq:QQNM_notation} is generated by four linear
    QNMs:
    \begin{align}
        (\ell_{1},m_{1},n_{1},p_{1}), &  & (\ell_{1},-m_{1},n_{1},-p_{1}), \notag \\
        (\ell_{2},m_{2},n_{2},p_{2}), &  & (\ell_{2},-m_{2},n_{2},-p_{2}), \notag
    \end{align}
    with respective strain amplitudes $A_{1}$, $A_{-1}$, $A_{2}$, and $A_{-2}$. It is
    worth noting that while the frequency values of
    $(\ell_{1},m_{1},n_{1},p_{1})$ and $(\ell_{1},-m_{1},n_{1},-p_{1})$ are
    related via Eq.~\eqref{eq:qnm_kerr_symmetry}, their amplitudes $A_{\pm 1}$ are independent, except in nonprecessing systems, where reflection symmetry imposes
    \begin{align}
        A_{-1}=(-1)^{\ell_1}\bar{A}_{1}. \label{eq:nonprecessing_linear_amplitudes}
    \end{align}
    The four linear QNMs, coupled through the following four channels give rise
    to the same QQNM frequency:
    \begin{subequations}
        \begin{align}
             & (\ell_{1},m_{1},n_{1},p_{1}) \otimes (\ell_{2},m_{2},n_{2},p_{2}),     &  & ++, \label{eq:++_channel} \\
             & (\ell_{1},m_{1},n_{1},p_{1}) \otimes (\ell_{2},-m_{2},n_{2},-p_{2}),   &  & +-,\label{eq:+-_channel}  \\
             & (\ell_{1},-m_{1},n_{1},-p_{1}) \otimes (\ell_{2},m_{2},n_{2},p_{2}),   &  & -+,                       \\
             & (\ell_{1},-m_{1},n_{1},-p_{1}) \otimes (\ell_{2},-m_{2},n_{2},-p_{2}), &  & --.
        \end{align}
    \end{subequations}
    Below we will refer to them as $++$, $+-$, $-+$, and $--$, respectively. For
    example, the $(+-)$ channel corresponds to the interaction where the linear mode
    $(\ell_{2},-m_{2},n_{2},-p_{2})$ generates a frequency component $-\bar{\omega}
    _{\ell_2-m_2n_2}^{-p_2}$ in the linearized metric, which is numerically
    equal to $\omega_{\ell_2m_2n_2}^{p_2}$ due to Eq.~\eqref{eq:qnm_kerr_symmetry}.
    This component then interacts with the frequency $\omega_{\ell_1m_1n_1}^{p_1}$
    generated by the first linear mode $(\ell_{1},m_{1},n_{1},p_{1})$. Denoting
    the strain amplitude of the QQNM
    as $A_{Q}$ (using either the spherical or spheroidal basis), we have
    \begin{align}
        A_{Q}= & R_{++}A_{1}A_{2}+R_{+-}A_{1}\bar{A}_{-2}+R_{-+}\bar{A}_{-1}A_{2}\notag \\
               & +R_{--}\bar{A}_{-1}\bar{A}_{-2}. \label{eq:AQ_Al}
    \end{align}
    where the coefficients $R$'s are fully determined by the dimensionless spin
    of a Kerr BH, and the angular mode under consideration. Here we use
    spin-weighted spherical/spheroidal harmonics as the mode basis to facilitate
    direct comparison with previous results, such as those from numerical
    relativity and from Paper I. Alternatively, modes can also be classified by
    parity, see the Supplementary Material for more details.

    For nonprecessing binaries, Eq.~\eqref{eq:AQ_Al} reduces to
    \begin{align}
        R^{\rm total}\equiv \frac{A_{Q}}{A_{1}A_{2}}= & R_{++}+(-1)^{\ell_2}R_{+-}\notag                                                       \\
                                                      & +(-1)^{\ell_1}R_{-+}+(-1)^{\ell_1+\ell_2}R_{--}, \label{eq:total_ratio_non_precessing}
    \end{align}
    where we have used Eq.~\eqref{eq:nonprecessing_linear_amplitudes}. Notice that
    the right-hand side (RHS) is an intrinsic property of a Kerr BH, independent
    of the linear mode amplitudes. We define the RHS as the total amplitude ratio
    $R^{\rm total}$, which characterizes the excitability of the QQNM.

    The $(--)$ channel deserves more attention. We find that $R_{--}$ always vanishes
    identically, regardless of the types of parent modes and the spin of the
    Kerr BH. This result follows directly from the second-order Teukolsky equation
    in Eq.~\eqref{eq:second_order_Teukolsky_eqn}, where the source $S_{4}^{(2)}$
    takes the following form in ORG \cite{Ma:2024qcv}:
    \begin{align}
        S_{4}^{(2)}= (\ldots)^{(1)}\Psi_{4}^{(1)}+ (\ldots)^{(1)}\Psi_{3}^{(1)}+ (\ldots)^{(1)}\Psi_{2}^{(1)}. \label{eq:source_S4}
    \end{align}
    The structure of $S_{4}^{(2)}$ explicitly illustrates how two parent modes
    couple. The first parent mode contributes the three Weyl scalars
    $\Psi_{4}^{(1)}$, $\Psi_{3}^{(1)}$, and $\Psi_{2}^{(1)}$, which are then multiplied
    by the terms contributed by the second parent mode (indicated in parentheses).
    For a given linear QNM $(l,m,n,p)$, the expressions for $\Psi_{4}^{(1)}$,
    $\Psi_{3}^{(1)}$, and $\Psi_{2}^{(1)}$ were analytically derived in Paper I \cite{Ma:2024qcv},
    see Eqs.~(38), (42), and (46) therein. An intriguing feature emerges: all three
    Weyl scalars contain only a single frequency component, $\omega_{lmn}^{p}$, with
    no contribution from the $-\bar{\omega}_{lmn}^{p}$ frequency component
    in the linearized metric that is associated with the mode $(\ell, m, n, p)$. As a result, the source in Eq.~\eqref{eq:source_S4}
    contains no terms resulting from the product of two conjugated frequency components.

    \section{Results and Comparisons}
    The QQNM $\begin{pmatrix}
        220+ \\
        220+
    \end{pmatrix}$ in the $l=m=4$ harmonic has been extensively discussed in the
    literature \cite{Ma:2022wpv,Mitman:2022qdl,Cheung:2022rbm,Zhu:2024rej,Redondo-Yuste:2023seq,Bucciotti:2024zyp,Bucciotti:2024jrv,Bourg:2024jme,Ma:2024qcv}.
    In particular, in Paper I, we found that the computed amplitude ratio, $R_{++}$ for the $(4,4)$ mode in Eq.~\eqref{eq:AQ_Al}, was
    smaller than those extracted from numerical relativity
    \cite{Zhu:2024rej,Redondo-Yuste:2023seq}. We speculated that the difference was
    due to parity freedom in the parent modes (see Sec.~VI.~B in
    \cite{Ma:2024qcv}). This was later confirmed in
    \cite{Bucciotti:2024zyp,Bucciotti:2024jrv,Bourg:2024jme} for Schwarzschild
    BHs. In this work, after classifying all mode-coupling channels, we revisit the
    discrepancy for Kerr BHs.

    \begin{figure}[!ht]
        \centering
        \includegraphics[width=\linewidth]{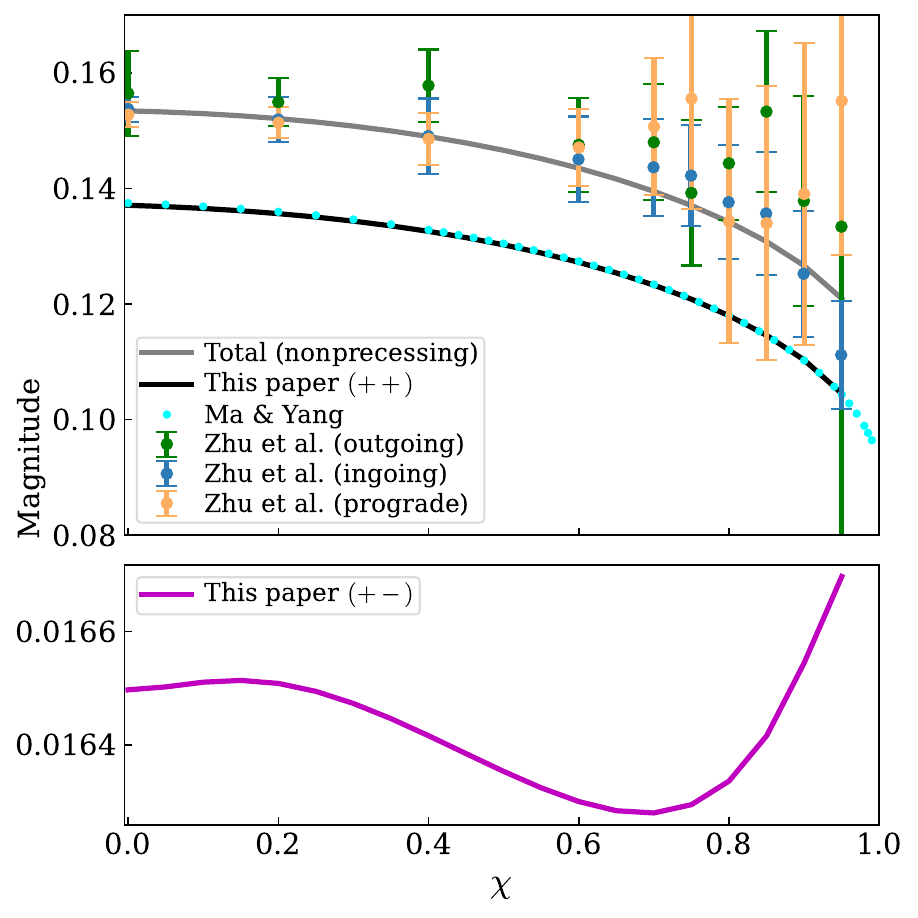}
        \includegraphics[width=\linewidth]{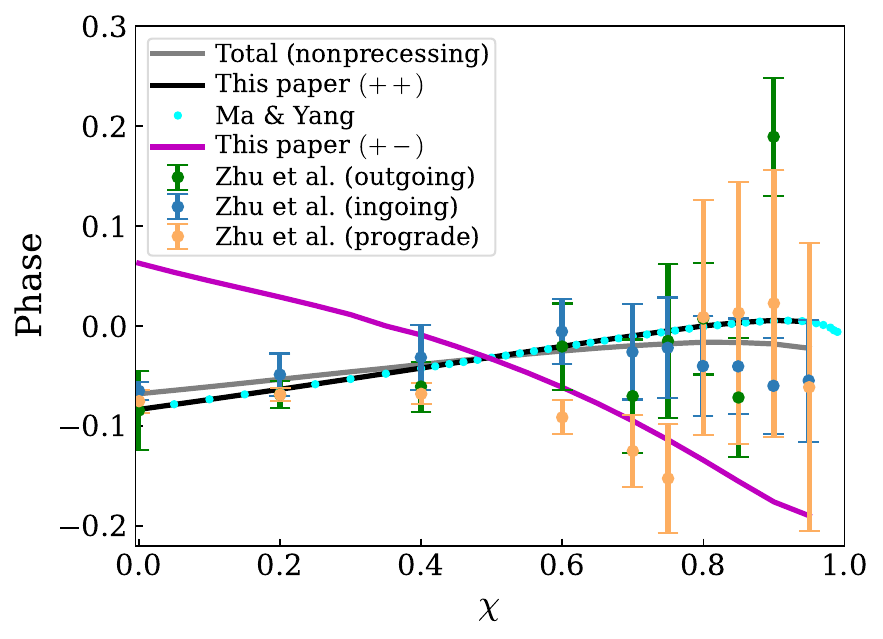}
        \caption{A comparison between the complex contour method (cyan dots,
        from Paper I), time-domain numerical simulations (green, blue, and orange
        dots, from Zhu \etal \cite{Zhu:2024rej}), and our current hyperboloidal
        slicing method (gray, black, and magenta curves). The gray curve represents
        the total amplitude ratio $R^{\rm total}$ for nonprecessing binaries [Eq.~\eqref{eq:total_ratio_non_precessing}];
        the black curve corresponds to $R_{++}$ in Eq.~\eqref{eq:++_channel}; and
        the magenta curve shows $R_{+-}$ in Eq.~\eqref{eq:+-_channel}. The mode
        amplitudes extracted by Zhu \etal~\cite{Zhu:2024rej} used spin-weighted
        spherical harmonics, while Paper I adopted spin-weighted spheroidal harmonics.
        Here we use spin-weighted spheroidal harmonics to make a fair comparison
        with Paper I. Results between these two angular bases differ by $\le 1\%$
        (see Supplementary Material).}
        \label{fig:ratio_comparison_2201_2201}
    \end{figure}

    As shown by the black solid lines in Fig.~\ref{fig:ratio_comparison_2201_2201},
    we first compute the magnitude and phase of $R_{++}$ as functions of the
    dimensionless spin $\chi$. Good agreement is observed with the results obtained
    in Paper I, as represented by cyan dots. The magenta curves correspond to the
    results for $R_{+-}$. Unlike $R_{++}$, the magnitude of $R_{+-}$ exhibits a weaker
    and non-monotonic dependence on spin, and its value is smaller than that of
    $R_{++}$ by a factor of $\sim7$.

    For nonprecessing binaries, the total excitability of the QQNM is characterized
    by $R^{\rm total}$ [Eq.~\eqref{eq:total_ratio_non_precessing}]. The gray
    curves in Fig.~\ref{fig:ratio_comparison_2201_2201} show its dependence on spin.
    For a Schwarzschild BH, $R^{\rm total}=0.1534e^{-0.06787i}$, in agreement
    with the values reported in
    \cite{Bucciotti:2024zyp,Bucciotti:2024jrv,Bourg:2024jme}. For Kerr BHs, the
    results (for $R^{\rm total}$) extracted from numerical-relativity scattering
    experiments \cite{Zhu:2024rej} are shown as green, blue, and orange dots in
    Fig.~\ref{fig:ratio_comparison_2201_2201}. Our findings are consistent with
    theirs.

    We note that the values of $R$'s depend on the choice of angular basis ---
    Paper I adopts spin-weighted spheroidal harmonics whereas numerical
    relativity simulations use spin-weighted spherical harmonics. In the Supplementary
    Material, we compare $R$ across these two bases, finding that they differ by
    only $\sim 1\%$ even when $\chi=0.95$. For a fair comparison with Paper I, we
    use spin-weighted spheroidal harmonics for the black and magenta curves in Fig.~\ref{fig:ratio_comparison_2201_2201}.

    Besides $\begin{pmatrix}
        220+ \\
        220+
    \end{pmatrix}$, we also study six more QQNMs, including $\begin{pmatrix}
        330+ \\
        220+
    \end{pmatrix}$, $\begin{pmatrix}
        440+         \\
        2\text{-}20-
    \end{pmatrix}$, $\begin{pmatrix}
        440+         \\
        3\text{-}20-
    \end{pmatrix}$,
    $\begin{pmatrix}
        220+ \\
        200+
    \end{pmatrix}$, $\begin{pmatrix}
        220+ \\
        200-
    \end{pmatrix}$, and $\begin{pmatrix}
        330+         \\
        2\text{-}20-
    \end{pmatrix}$. These modes are heuristically picked, motivated by the observational relevance of the $(3,3,0,+)$ mode\footnote{Other linear modes, such as $(210+)$ \cite{Siegel:2023lxl,Zhu:2023fnf}, are also relevant.} and the the recent QQNMs discovered in numerical waveforms~\cite{Giesler:2024hcr}.
    We use polynomials to fit their $R$ for
    different channels as functions of $\chi$:
    \begin{align}
        R=Ae^{i\phi}, \quad \text{with}~A=\sum_{i=0}^{7}A_{i}\chi^{i},\phi=\sum_{i=0}^{7}\phi_{i}\chi^{i}. \label{eq:fitting_R_chi}
    \end{align}
    The values of the coefficients are provided in tables in the Supplementary Material. 

    \section{Implications for future ringdown data analysis}
    
    We now analyze the significance of the QQNMs on future ringdown data analysis by including all channels. For a binary system, the remnant surrogate model from \cite{Varma:2018aht} predicts the remnant's spin, which can be used in Eq.~\eqref{eq:fitting_R_chi} to compute quadratic amplitude ratios. When combined with the \texttt{postmerger} ringdown model \cite{Pacilio:2024tdl}, these ratios enable precise quadratic amplitude predictions. This information has the potential to improve future Bayesian analyses by narrowing prior ranges and reducing the parameter space. Note that our calculations include precessing systems.

    Designing a systematic Bayesian framework that incorporates our results is beyond the scope of this paper. Instead, as a figure of merit, we illustrate a practical application of our calculations by estimating the SNR of various QQNMs that have not been previously discussed in the literature. Below, we follow the method outlined in \cite{Flanagan:1997sx,Berti:2005ys,Yi:2024elj}
    to compute these SNRs. For simplicity, we average the SNRs over all angles, which
    yields the orthogonal condition between two antenna patterns $F_{+,\times}$ for
    CE:
    \begin{align}
        \left<F_{+,\times}^{2}\right>=\frac{1}{5}, \quad \left<F_{+}F_{\times}\right>=0. \notag
    \end{align}
    For LISA, there is an additional factor of $3/2$. As pointed out in Paper I,
    a QQNM is present in different $\ell$ harmonics. Its total SNR is given by
    \begin{align}
        \rho = \sqrt{\sum_{\ell}\rho_{\ell,m}^{2}}.
    \end{align}
    In practice, we find that the first $\ell$ harmonic typically dominates the total
    SNR.

    \begin{figure}[tb]
        \centering
        {\includegraphics[width=\linewidth,clip=true]{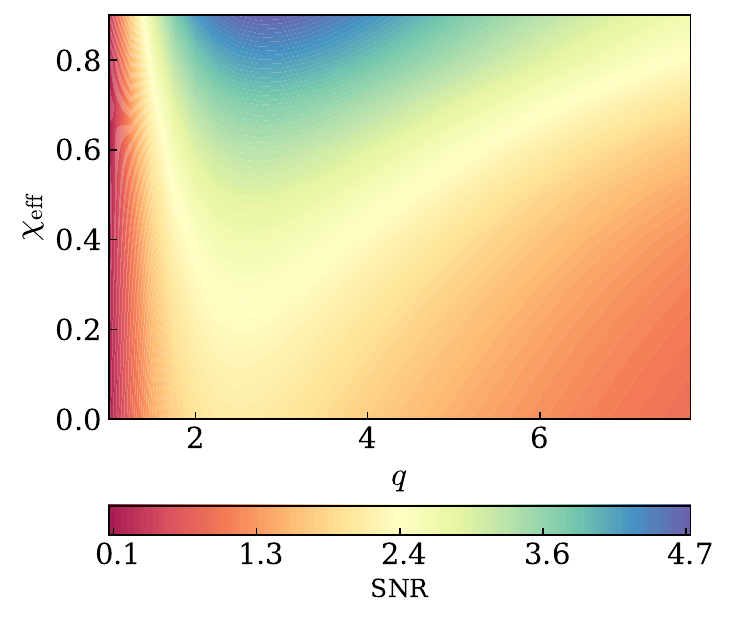}}
        \caption{The SNR dependence of the QQNM, $\begin{pmatrix}
            3, & 3, & 0, & + \\
            2, & 2, & 0, & +
        \end{pmatrix}$, on progenitor binaries' mass ratio $q$ and effective
        spin $\chi_{\rm eff}$, assuming CE. The binary systems have a total mass of ${60}M_{\odot}$ and
        a luminosity distance of 410 Mpc.}
        \label{fig:330+_220+_SNR_2D}
    \end{figure}

    In this paper, we adopt a conservative estimate by choosing the starting time
    to be $10M_{\rm tot}$ after the peak, where $M_{\rm tot}$ is the total mass
    of the binary system. Selecting an earlier starting time can moderately
    increase the SNRs, since QQNMs decay faster than linear fundamental modes.

    Figure \ref{fig:330+_220+_SNR_2D} shows the SNR dependence of the QQNM $\begin{pmatrix}
        330+ \\
        220+
    \end{pmatrix}$ on the mass ratio $q$ and the effective inspiral
    spin $\chi_{\rm eff}$. This analysis is performed for CE using $M_{\rm tot}=60M_\odot$ and a luminosity distance of 410 Mpc. For simplicity, we set both component spins equal, $\chi_{1}=\chi_{2}=\chi_{\rm eff}$. We can see that the optimal binary configuration occurs at $q\sim 2.75$ and maximally positive $\chi_{\rm eff}$, yielding SNR $\sim4.7$.

     Here, we focus on a representative binary by fixing
    $M_{\rm tot}$ and $D_{L}$. The total mass affects the SNR in two ways: (1)
    it scales the overall strain linearly; and (2) it sets the central frequency
    of a QQNM. Varying $M_{\rm tot}\in [20,200]M_{\odot}$, we find that SNR increases monotonically from 0.2 to 59. The corresponding plot is included in the Supplementary Material.

    Besides $\begin{pmatrix}
        330+ \\
        220+
    \end{pmatrix}$, Table \ref{table:QQNMs_SNR} summarizes the SNRs of various QQNMs. Most are significant enough for LISA detection, while for CE, the SNRs fall just below the typical threshold ($\sim 8$) at the given $M_{\rm tot}$ and $D_{L}$, though they could be higher for more massive or closer binaries. Nevertheless, our results suggest the importance of subdominant QQNMs and the difficulty of future ringdown data analysis:
    While the QQNMs may lack sufficient SNR for precise parameter estimation, omitting them --- as noted in \cite{Giesler:2024hcr} --- can introduce systematic errors in linear mode fitting, destabilizing extracted amplitudes. Thus, a robust ringdown analysis pipeline will be essential for future detectors. 

  \begin{table}
    \centering
    \caption{Optimal SNRs of various QQNMs, assuming CE with $M_{\rm tot}=60M_\odot$ and $D_L=410$ Mpc, and LISA with $M_{\rm tot}=8\times 10^6M_\odot$ and a redshift of 1. The SNRs are obtained from optimal binary configurations that maximize SNR for the specified $M_{\rm tot}$ and $D_L$.}
    \begin{tabular}{c c c c} \hline\hline
     \multirow{2}{*}{QQNM} & Optimal  &  \multicolumn{2}{c}{SNR}  \\
     & configuration & CE & LISA  \\ \hline
     $\begin{pmatrix}
            2, & 2, & 0, & + \\
            2, & 2, & 0, & +
        \end{pmatrix}$ &\makecell{$q\sim1$ \\ large $\chi_{\rm eff}$} & 9 & 970 \\  \hline
     $\begin{pmatrix}
            3, & 3, & 0, & + \\
            2, & 2, & 0, & +
        \end{pmatrix}$ &\makecell{$q\sim 2.75$ \\ large $\chi_{\rm eff}$} & 4.7 & 653 \\ \hline 
    $\begin{pmatrix}
            4, & 4, & 0, & + \\
            2, & -2, & 0, & -
        \end{pmatrix}$ &\makecell{$q\sim 4.25$ \\ large $\chi_{\rm eff}$} & 3 & 140  \\ \hline
    $\begin{pmatrix}
            3, & 3, & 0, & + \\
            2, & -2, & 0, & -
        \end{pmatrix}$ &\makecell{$q\sim 2.75$ \\ large $\chi_{\rm eff}$} & 5 & 60  \\ \hline 
    $\begin{pmatrix}
            4, & 4, & 0, & + \\
            3, & -2, & 0, & -
        \end{pmatrix}$ &\makecell{$q\sim 8$ \\ large $\chi_{\rm eff}$} &$0.084$  & 2.1  \\ \hline \hline
     \end{tabular}
     \label{table:QQNMs_SNR}
\end{table}

    \section{Discussion}
    Together with Paper I, we have presented two independent frequency-domain methods
    for analyzing QQNMs --- the complex contour technique and the hyperboloidal slicing
    approach. Both methods produce consistent results for $R_{++}$ of
    $\begin{pmatrix}
        220+ \\
        220+
    \end{pmatrix}$. We have also classified the four possible coupling channels for
    a QQNM. Notably, the coupling between two mirrored modes (labeled by $--$) vanishes
    exactly. After accounting for all coupling channels in systems with
    reflection symmetry, our results for Kerr BHs match those from numerical simulations
    and time-domain fits, which therefore resolves the tension between Paper I and Zhu \etal \cite{Zhu:2024rej}.

    The agreement between our theoretical predictions and numerical relativity results provides a compelling quantitative demonstration---one of only a few such examples (e.g. \cite{Nicasio:1998aj,Gleiser:1996yc})---of the success of BH second-order perturbation theory. Our theoretical analysis reveals that the quadratic ratios reported in \cite{Zhu:2024rej,Redondo-Yuste:2023seq} are limited to nonprecessing binaries. By contrast, the excitation of QQNMs in precessing binaries is expected to encode richer information about parent modes and progenitors' parity. This opens an avenue for future studies to investigate how QQNMs amplitudes can infer precession parameters.

    We have also illustrated potential applications of our calculations in future ringdown data analysis. As a figure of merit, we performed an SNR survey. Our results indicate that multiple QQNMs are likely to play a nonnegligible role in BH spectroscopy with future detectors. Therefore, a robust ringdown data analysis framework will be required to account for a substantial number of linear modes and QQNMs, in order to avoid systematic biases and prevent false alarms of deviations from GR.

    Although this work, together with Paper I, has offered a comprehensive
    discussion on QQNMs, a complete account of all second-order effects has yet to
    be provided. Intriguing avenues for future study include the memory effect
    of a QNM, quadratic tails, and the possibility of resolving the nonlinear
    tails from numerical simulations, using techniques such as Cauchy-characteristic
    matching \cite{Ma:2023qjn}.


    \begin{acknowledgments}
        \section{Acknowledgments}
        We thank Hengrui Zhu for sharing the data in Fig.~\ref{fig:ratio_comparison_2201_2201}.
        We thank Andrew Spiers for pointing out the difference between the Campanelli–Lousto–Teukolsky
        equation and the reduced second-order Teukolsky equation. Research at
        Perimeter Institute is supported in part by the Government of Canada
        through the Department of Innovation, Science and Economic Development and
        by the Province of Ontario through the Ministry of Colleges and Universities.
        N.K. is supported by the Natural Science and Engineering Council of Canada.
    \end{acknowledgments}


    \bibliography{references}

    \section{SUPPLEMENTARY MATERIAL}
\setcounter{page}{1}

\subsection{Description of the numerical code}

As discussed before, we solve the nonlinear Teukolsky equations on a hyperboloidal
slicing. We use identical coordinates as in~\cite{Ripley:2022ypi}. Namely, for
the Kerr metric
\begin{align}
    \label{eq:Kerr}\rmd s^{2} & = - \left( 1- \frac{2Mr}{\Sigma}\right) \rmd t^{2}- \frac{4Mar\sin^{2}\theta}{\Sigma}\, \rmd t\,\rmd\phi + \frac{\Sigma}{\Delta}\rmd r^{2}\nonumber \\
                              & + \Sigma \,\rmd\theta^{2}+ \left(r^{2}+a^{2}+ \frac{2M a^{2}r\sin^{2}\theta}{\Sigma}\right) \sin^{2}\theta\,\rmd\phi^{2}\,,
\end{align}
where $\Sigma = r^{2}+ a^{2}\cos^{2}\theta$ and $\Delta= r^{2}-2Mr+a^{2}$, we change
the coordinates $(t, r, \phi)$ to $(T,R, \Phi)$ where
\begin{subequations}
    \begin{align}
        \rmd T    & = \rmd t + \frac{r^{2}+ a^{2}}{\Delta}dr - 2\left(1+\frac{2M}{r}\right)\rmd r , \\
        \rmd \Phi & = \rmd\phi + \frac{a \rmd r}{\Delta},                                           \\
        R         & = \frac{1}{r}\,,
    \end{align}
\end{subequations}
and where the coordinate $\theta$ remains unchanged. Note that the coordinates
$T$ and $\Phi$ are chosen such that they are well-defined at the horizon. Furthermore,
$T=\mathrm{const}$ slices are hyperboloidal, namely they penetrate the horizon and
null infinity, and become light-like in the $r\to\infty$ limit. The coordinate $R$
is chosen to compactify the space-time radially onto a finite domain.

The principal null tetrad must also be chosen to be regular at both the horizon and
null-infinity. While the Kinnersley tetrad is regular at null-infinity, it is not
a good choice at the horizon. Therefore, we choose the tetrad $(\ell^{a}, n^{a},
m^{a}, \bar{m}^{a})$ given by
\begin{align}
    \ell^{a} & = \frac{\Delta}{2r^{2}}\ell^{a}_{K}, \\
    n^{a}    & = \frac{2r^{2}}{\Delta}n^{a}_{K},    \\
    m^{a}    & = m^{a}_{K},
\end{align}
where $(\ell^{a}_{K}, n^{a}_{K}, m^{a}_{K}, \bar{m}^{a}_{K})$ is the Kinnersley null
tetrad.

Under these coordinates and conventions the spin-$s$ radial Teukolsky equation for
the spin-$s$ field $\psi_{s}$, with $\psi_{s}= R^{1+2s}\psi^{\circ}_{s}$ becomes
\begin{equation}
    \label{eq:Teukolsky-radial}\alpha \psi^{\circ}_{s}+ \beta \partial_{R}\psi^{\circ}
    _{s}-(1-2MR^{2}+a^{2}R^{2})R^{2}\partial_{R}^{2}\psi^{\circ}_{s}= - (\Lambda
    /2+1)\psi^{\circ}_{s}\,,
\end{equation}
where $\Lambda$ is the separation constant of the angular Teukolsky equation and
$\alpha$ and $\beta$ are given by
\begin{align}
    \alpha & = a^{2}R^{2}\left(8 M^{2}\omega^{2}+6 i M \omega -1\right)\nonumber                                      \\
           & -R (4 M \omega +i) \left(-a (a \omega +m)+4 M^{2}\omega +i M (s+1)\right)\nonumber                       \\
           & +\frac{1}{2}\omega^{2}\left(a^{2}-16 M^{2}\right)+a m \omega +s \left(-\frac{1}{2}+2 i M \omega \right), \\
    \beta  & = 2 a^{2}R^{3}(-1+2 i M \omega )+R (-s-1)+i \omega\nonumber                                              \\
           & +R^{2}\left(i a (a \omega +m)-8 i M^{2}\omega +M (s+3)\right)\,.
\end{align}
For instance, for the Teukolsky equation of $\Psi_{4}$ we have $s=-2$ and $\psi_{s}
= \zeta^{4}\Psi_{4}$, where $\zeta=r=ia\cos\theta$. Whereas for $\Psi_{0}$ we have
$s=2$ and $\psi_{s}= \Psi_{0}$.

To solve the second-order Teukolsky equation, we need the reconstructed metric
at linear order to compute the source term. We obtain the metric by first solving
for the ORG Hertz potential and then performing metric reconstruction. We use
the ORG because the metric reconstructed from the ingoing radiation gauge Hertz
potential is not well-behaved at null infinity. To solve the Hertz potential $\Phi$
we note that $\zeta^{-4}\Phi$ satisfies the $s=2$ Teukolsky equation, where
$\zeta = r-ia\cos\theta$. Consequently, we need to solve the $s=2$ radial
Teukolsky equation to get a solution for the Hertz potential that corresponds to
QNMs for the reconstructed metric.

The metric, reconstructed from the Hertz potential, in GHP notation~\cite{Geroch:1973am},
is given by
\begin{align}
    \label{eq:reconstruction}h_{ab}= - \frac{1}{2}\Big[ & n_{a}n_{b}(\eth' - \tau')(\eth' - \tau') \nonumber                         \\
                                                        & + \bar{m}_{a}\bar{m}_{b}(\thorn'-\rho')(\thorn'- \rho')\nonumber           \\
                                                        & - n_{(a}\bar{m}_{b)}\Big((\eth'-\tau'+\bar{\tau})(\thorn'-\rho') \nonumber \\
                                                        & \;+ (\thorn'-\rho'+\bar{\rho}')(\eth'-\tau')\Big) \Big]\Phi + c.c.
\end{align}
Evidently from~\eqref{eq:reconstruction} the gravitational strain
$h = h_{ab}\bar{m}^{a}\bar{m}^{b}$ (and therefore $\Psi_{4}$) only has
contributions from $\bar{\Phi}$, instead of $\Phi$ directly. Consequently, to get
a QNM frequency $\omega_{\ell m n}^{p}$ in $h$, we must have the frequency
$-\bar{\omega}_{\ell m n}^{p}$ and azimuthal quantum number $-m$ in the Hertz
potential. Note however that the outgoing and ingoing boundary conditions do not
get mixed under complex conjugation or derivatives, therefore if the strain $h$
satisfies purely outgoing boundary conditions, so does the Hertz potential $\Phi$.
Therefore, we find the QNM solutions for the mode $(\ell, m, n, p)$ by finding outgoing
solutions to the Hertz potential $\Phi$ with azimuthal quantum number $-m$ and frequency
$-\bar{\omega}_{\ell m n}^{p}$.

Our numerical implementation uses the pseudo-spectral method, where all
quantities are expanded in Chebyshev polynomials radially and spin-weighted
spherical harmonics in the angular directions. Multiplications are performed
using fast transforms from the spectral basis to a grid. All results in this
paper have radial expansion truncated at degree $n_{\text{max}}=64$, the spherical
harmonic expansion truncating at $\ell_{\text{max}}=31$ and with 128 bit floating
point numbers from the \texttt{MultiFloats} library~\cite{MultiFloats}.
    
We solve the equation for the Hertz potential following the procedure in~\cite{Ripley:2022ypi},
which we summarize below. We first separate the Hertz potential into radial and angular
parts
\begin{equation}
    \Phi = R\,\mathcal{R}(R)\,\mathcal{S}(\theta, \Phi) e^{- i\omega T}\,,
\end{equation}
which can be done because of the separation of the Teukolsky equation. Note we
have also separated the leading order fall off at null infinity $\sim R$ so that
$\mathcal{R}$ is $\mathcal{O}(1)$ at null infinity. Then, the radial and angular
functions $\mathcal{R}$ and $S$ can be expanded into Chebyshev polynomials
$T_{n}(x)$ and the spin-weighted spherical harmonics
${}_{2}Y_{\ell m}(\theta, \phi)$ respectively.
\begin{align}
    \mathcal{R}(R)           & = \sum_{n=0}^{\infty}c_{n}T_{n}(1-2Rr_{+}),                \\
    \mathcal{S}(\theta,\Phi) & = \sum_{\ell}s_{\ell}\;\,{}_{2}Y_{\ell m}(\theta, \Phi)\,.
\end{align}
The outgoing boundary conditions at the horizon and null infinity imply that $\mathcal{R}$
is analytic~\cite{Ripley:2022ypi} at the boundary and thus $c_{n}$ falls exponentially
with $n$, whereas for the ingoing boundary conditions, the $c_{n}$ falls faster than
any polynomial because the solution is smooth, yet it falls slower than an
exponential. Thus, just as in Leaver's method the QNM solutions are the minimal
solution of the radial equation. Then, if we impose the conditions
$c_{n_{\text{max}+1}}=c_{n_{\text{max}+2}}=0$ for a large enough
$n_{\text{max}}$ and solve for the truncated coefficients
$c_{0}\ldots c_{n_{\text{max}}}$, the solution will be dominated by outgoing solutions
just as for Leaver's method (if a non-zero solution exists for that frequency).
Additionally, we can also truncate the angular expansion at some $\ell_{\text{max}}$
for numerical purposes.

The truncated Radial and Angular Teukolsky equations now turn into matrix eigenvalue
problems, with the separation constant $\Lambda$ related to their respective
eigenvalues. We change the basis after differentiating to Gegenbauer polynomials
to improve the conditioning of the linear problem~\cite{OlverTownsend}. For
generic frequencies, the two equations will not give the same value of $\Lambda$,
however for QNMs they do. Therefore, following~\cite{Ripley:2022ypi}, we find QNMs
by minimizing the difference between the separation constants obtained from the radial
and angular equations, using the frequencies from the \texttt{qnm} package as a good
initial guess~\cite{Stein:2019mop}. This gives us the Hertz potential as a
series in Chebyshev and spin-weighted spherical harmonics. We do this for both pairs
of linear modes with frequencies $\omega_{1}$ and $\omega_{2}$ we want to compute
the quadratic contribution to and take their linear combination.

From the Hertz potential, we use~\eqref{eq:reconstruction} to obtain the linearized
metric. Here derivatives become linear operators that can be applied to the
spectral coefficients. To perform multiplication, we first transform the
spectral data into a grid of 2d collocation points by performing the fast
discrete cosine transform on the radial direction, and an adaptation of the fast
spin-weighted spherical harmonics transform described in~\cite{Huffenberger_2010}
for the angular direction, where the grid in the $\phi$ direction is not
computed because our fields have a fixed azimuthal quantum number $m$. Then after
point-wise multiplication on the grid, we transform back to the spectral domain
using the inverse fast discrete cosine transform and the inverse fast spin-weighted
spherical harmonics transform. We keep track of the fall-off in $R$ of all the terms
and perform spectral expansion of the $\mathcal{O}(1)$ parts.

Finally, using the metric, we compute the Einstein tensor $G_{ab}^{(2)}[h_{ab}, h
_{ab}]$ quadratic in $h_{ab}$. Here again, we perform multiplications by transforming
to a 2d grid of points. As discussed in the main text, $T_{ab}= -G_{ab}^{(2)}[h_{ab}
, h_{ab}]/8\pi$ acts as the effective stress-energy tensor for the next order.
The Teukolsky equation derived from this equation is the reduced Teukolsky
equation~\cite{soton469806}, because it only gives the second order $\Psi_{4}$ computed
linearly from $h_{ab}^{(2)}$ and does not include contributions from the quadratic
terms in $h_{ab}^{(1)}$. However, it can be checked that in ORG this contribution
vanishes, and therefore the reduced Teukoslky equation gives the full $\Psi_{4}$.
From this effective stress-energy tensor we can compute the source term $S$ for the
$s=-2$ reduced Teukolsky equation using
\begin{align}
    S & = \tfrac12(\eth'-\bar{\tau}-4\tau') \big[(\thorn'-2\bar{\rho}') T_{n\bar{m}}-(\eth'-\bar{\tau}) T_{nn}\big] \nonumber        \\
      & + \tfrac12(\thorn'-4\rho'-\bar{\rho}') \big[(\eth'-2\bar{\tau}) T_{n\bar{m}}- (\thorn'-\bar{\rho}') T_{\bar{m}\bar{m}}\big],
\end{align}
and then solving the sourced Teukolsky equation
\begin{equation}
    \mathcal{T}[\Psi_{4}] = S.
\end{equation}
The resulting source term has frequencies $\omega_{1}+\omega_{2}$,
$\omega_{1}- \bar{\omega}_{2}$ and $-\bar{\omega}_{1}+ \omega_{2}$. As discussed
in the main text, the contribution to the $-\bar{\omega}_{1}- \bar{\omega}_{2}$
vanishes. This gives rise to three frequencies $\omega_{\text{quad}}$ in the second
order $\Psi_{4}$, and we solve each separately. This could be done by simply getting
the Teukolsky operator in matrix form on the spectral coefficients and inverting
it. However, we can utilize the separation of the Teukolsky equation to invert it
directly. We first re-expand the source term in a basis of Spheroidal harmonics
$_{-2}S_{\ell m_{\text{quad}}}(\theta, \phi; a\omega_{\text{quad}})$ using the
procedure described in~\cite{Ma:2024qcv}. Then, the coefficients of the
spheroidal harmonics satisfy the radial Teukolsky equation~\eqref{eq:Teukolsky-radial}
with a source. This reduces the dimensions of the problem, leading to faster computations.
After separating the leading order fall-off in $R$, the radial operator in~\eqref{eq:Teukolsky-radial}
can be expressed as a lower dimensional matrix. Then, the spheroidal harmonic
component of $S$ gives us the source term for the matrix equation. Finally, by
solving the matrix equation we get a solution to the quadratic $\Psi_{4}$ from
the horizon to null infinity. We can similarly solve for the reduced $\Psi_{0}$ as
well. However for $\Psi_{0}$ the quadratic terms from the $h_{ab}^{(1)}$ can
contribute. This can be added to the reduced $\Psi_{0}$ by using the $h_{ab}^{(1)}$
from the metric reconstruction. In this paper, we only report the reduced $\Psi_{0}$
which depends linearly on $h_{ab}^{(2)}$.

\subsection{Coupling strength for selected quadratic channels}
We compute the amplitude ratio $R$ for five QQNMs, including $\begin{pmatrix}
    220+ \\
    220+
\end{pmatrix}$, $\begin{pmatrix}
    330+ \\
    220+
\end{pmatrix}$, $\begin{pmatrix}
    440+         \\
    2\text{-}20-
\end{pmatrix}$, $\begin{pmatrix}
    220+ \\
    200+
\end{pmatrix}$, and $\begin{pmatrix}
    330+         \\
    2\text{-}20-
\end{pmatrix}$. We then fit their magnitudes and phases using polynomials as functions
of $\chi$:
\begin{align}
    R=Ae^{i\phi}, \quad \text{with}~A=\sum_{i=0}^{7}A_{i}\chi^{i},\,\phi=\sum_{i=0}^{7}\phi_{i}\chi^{i}.
\end{align}
The values of $A_{i}$'s and $\phi_{i}$'s are listed in Table \ref{tab:abs_fits}
and \ref{tab:ph_fits}, respectively. Here Spin-weighted spherical harmonics are used as the angular basis. { We estimate the relative interpolation errors, by computing the leave-one-out root mean squared error, to be of order $\sim 1\%$ for the phase and amplitude. This is much larger than the errors in computation of the quadratic ratio, which is of order $10^{-5}$ (this is estimated by comparison of our results with a lower resolution run using $n_{max}=60, \ell_{max}=27$ and using floating point precision). Thus the error is dominated by the interpolation error. }

\begin{table*}
    []
    \centering
    \scalebox{0.94}{
    \begin{tabular}{ccccccccccc}\toprule
\multirow{2}{*}{QQNM}  & Child Harmonic & \multirow{2}{*}{Channel} &\multirow{2}{*}{$ A_{0}  $} & \multirow{2}{*}{$ A_{1}  $} & \multirow{2}{*}{$ A_{2}  $} & \multirow{2}{*}{$ A_{3}  $} & \multirow{2}{*}{$ A_{4}  $} & \multirow{2}{*}{$ A_{5}  $} & \multirow{2}{*}{$ A_{6}  $} & \multirow{2}{*}{$ A_{7}  $} \\
&$(\ell, \, m)$& & & & & & & & & \\ \midrule
\multirow{ 2 }{*} { $\begin{pmatrix} 2 & 2 & 0 & + \\ 2 & 2 & 0 & + \end{pmatrix}$ }& \multirow{ 2 }{*}{ (4, 4) } & $(+,+)$ &0.137 & -0.00485 & 0.0167 & -0.2604 & 0.8975 & -1.617 & 1.434 & -0.5063 \\
 & & $(+,-)$ &0.01651 & 0.0004883 & -9.269e-05 & -0.02175 & 0.08652 & -0.1513 & 0.1265 & -0.04002 \\
\midrule\multirow{ 3 }{*} { $\begin{pmatrix} 3 & 3 & 0 & + \\ 2 & 2 & 0 & + \end{pmatrix}$ }& \multirow{ 3 }{*}{ (5, 5) } & $(+,+)$ &0.3608 & 0.02946 & 0.072 & -0.8283 & 2.78 & -4.998 & 4.432 & -1.568 \\
 & & $(+,-)$ &0.05032 & 0.003027 & 0.01122 & -0.1226 & 0.4119 & -0.7303 & 0.6395 & -0.2213 \\
 & & $(-,+)$ &0.009195 & 0.001647 & -0.0009933 & 0.005611 & -0.02125 & 0.04302 & -0.04239 & 0.01705 \\
\midrule\multirow{ 3 }{*} { $\begin{pmatrix} 4 & 4 & 0 & + \\ 2 & -2 & 0 & - \end{pmatrix}$ }& \multirow{ 3 }{*}{ (2, 2) } & $(+,+)$ &0.02926 & 0.01946 & -0.01687 & -0.1284 & 0.5675 & -0.9904 & 0.8196 & -0.2654 \\
 & & $(+,-)$ &0.4171 & 0.01547 & 0.05337 & -0.5621 & 1.897 & -3.389 & 2.988 & -1.045 \\
 & & $(-,+)$ &0.01904 & 0.0209 & -0.06647 & 0.1115 & -0.02876 & -0.1313 & 0.159 & -0.05317 \\
\midrule\multirow{ 3 }{*} { $\begin{pmatrix} 2 & 2 & 0 & + \\ 2 & 0 & 0 & + \end{pmatrix}$ }& \multirow{ 3 }{*}{ (2, 2) } & $(+,+)$ &0.05129 & 0.002671 & -0.03311 & 0.2839 & -0.9785 & 1.75 & -1.55 & 0.5454 \\
 & & $(+,-)$ &0.002266 & 0.007195 & -0.04674 & 0.377 & -1.297 & 2.319 & -2.053 & 0.7225 \\
 & & $(-,+)$ &0.002277 & -0.002354 & 0.0004335 & -0.002256 & 0.007851 & -0.01427 & 0.01362 & -0.005197 \\
\midrule\multirow{ 3 }{*} { $\begin{pmatrix} 3 & 3 & 0 & + \\ 2 & -2 & 0 & - \end{pmatrix}$ }& \multirow{ 3 }{*}{ (2, 1) } & $(+,+)$ &1.169 & -9.691 & 38.74 & -95.14 & 151.0 & -150.0 & 83.97 & -20.08 \\
 & & $(+,-)$ &0.1202 & -0.04895 & -0.3357 & 2.418 & -8.312 & 14.85 & -12.95 & 4.503 \\
 & & $(-,+)$ &1.06 & -8.979 & 34.85 & -76.17 & 102.4 & -84.76 & 39.9 & -8.12 \\
\midrule\bottomrule \end{tabular}
    }
    \caption{Polynomial fit coefficients for the magnitude of $R$. Spin-weighted
    spherical harmonics are used as the angular basis. When both modes are identical, $R_{+-}$ and $R_{-+}$ are equivalent, therefore only on of them is presented.
    }
    \label{tab:abs_fits}
\end{table*}
\begin{table*}
    \centering
    \scalebox{0.94}{
    \begin{tabular}{ccccccccccc}\toprule
\multirow{2}{*}{QQNM}  & Child Harmonic & \multirow{2}{*}{Channel} &\multirow{2}{*}{$ \phi_{0}  $} & \multirow{2}{*}{$ \phi_{1}  $} & \multirow{2}{*}{$ \phi_{2}  $} & \multirow{2}{*}{$ \phi_{3}  $} & \multirow{2}{*}{$ \phi_{4}  $} & \multirow{2}{*}{$ \phi_{5}  $} & \multirow{2}{*}{$ \phi_{6}  $} & \multirow{2}{*}{$ \phi_{7}  $} \\
&$(\ell, \, m)$& & & & & & & & & \\ \midrule
\multirow{ 2 }{*} { $\begin{pmatrix} 2 & 2 & 0 & + \\ 2 & 2 & 0 & + \end{pmatrix}$ }& \multirow{ 2 }{*}{ (4, 4) } & $(+,+)$ &-0.08375 & 0.09052 & 0.1573 & -0.947 & 3.078 & -5.277 & 4.554 & -1.574 \\
 & & $(+,-)$ &0.05858 & -0.1377 & -0.1421 & 1.362 & -6.616 & 13.41 & -12.76 & 4.635 \\
\midrule\multirow{ 3 }{*} { $\begin{pmatrix} 3 & 3 & 0 & + \\ 2 & 2 & 0 & + \end{pmatrix}$ }& \multirow{ 3 }{*}{ (5, 5) } & $(+,+)$ &-0.1163 & 0.09648 & 0.1244 & -0.7194 & 2.51 & -4.475 & 3.956 & -1.391 \\
 & & $(+,-)$ &0.1351 & -0.04493 & -0.1367 & 0.7989 & -2.826 & 4.871 & -4.206 & 1.394 \\
 & & $(-,+)$ &-3.006 & -0.1175 & -0.05493 & 0.1584 & -1.219 & 2.469 & -2.381 & 0.8356 \\
\midrule\multirow{ 3 }{*} { $\begin{pmatrix} 4 & 4 & 0 & + \\ 2 & -2 & 0 & - \end{pmatrix}$ }& \multirow{ 3 }{*}{ (2, 2) } & $(+,+)$ &-2.938 & -0.1743 & 3.982 & -16.57 & 40.29 & -57.42 & 44.05 & -14.09 \\
 & & $(+,-)$ &-2.924 & -0.08402 & 0.2896 & -1.953 & 6.776 & -12.12 & 10.75 & -3.79 \\
 & & $(-,+)$ &3.07 & -0.181 & 3.172 & -13.67 & 29.83 & -34.78 & 21.05 & -5.121 \\
\midrule\multirow{ 3 }{*} { $\begin{pmatrix} 2 & 2 & 0 & + \\ 2 & 0 & 0 & + \end{pmatrix}$ }& \multirow{ 3 }{*}{ (2, 2) } & $(+,+)$ &1.391 & 0.04208 & -0.005894 & 0.08979 & -0.3053 & 0.5355 & -0.4593 & 0.1514 \\
 & & $(+,-)$ &0.5606 & 0.2811 & -0.5344 & 4.077 & -13.84 & 24.83 & -22.05 & 7.8 \\
 & & $(-,+)$ &0.5613 & -0.2699 & 2.862 & -21.11 & 71.12 & -124.0 & 106.2 & -35.7 \\
\midrule\multirow{ 3 }{*} { $\begin{pmatrix} 3 & 3 & 0 & + \\ 2 & -2 & 0 & - \end{pmatrix}$ }& \multirow{ 3 }{*}{ (2, 1) } & $(+,+)$ &-2.696 & 2.556 & 1.259 & -19.84 & 96.27 & -177.5 & 133.6 & -33.46 \\
 & & $(+,-)$ &2.993 & -0.4982 & 0.3458 & -6.872 & 22.84 & -45.46 & 42.8 & -14.58 \\
 & & $(-,+)$ &0.6533 & 4.995 & -42.53 & 330.4 & -1002.0 & 1434.0 & -990.4 & 267.1 \\
\midrule\bottomrule \end{tabular}
    }
    \caption{ Polynomial fit coefficients for the phase of $R$. Spin-weighted
    spherical harmonics are used as the angular basis. When both modes are identical, the $+-$ and $-+$ channels are equivalent, therefore only on of them is presented.}
    \label{tab:ph_fits}
\end{table*}

\begin{figure}[!h]
    \includegraphics[width=\linewidth]{
        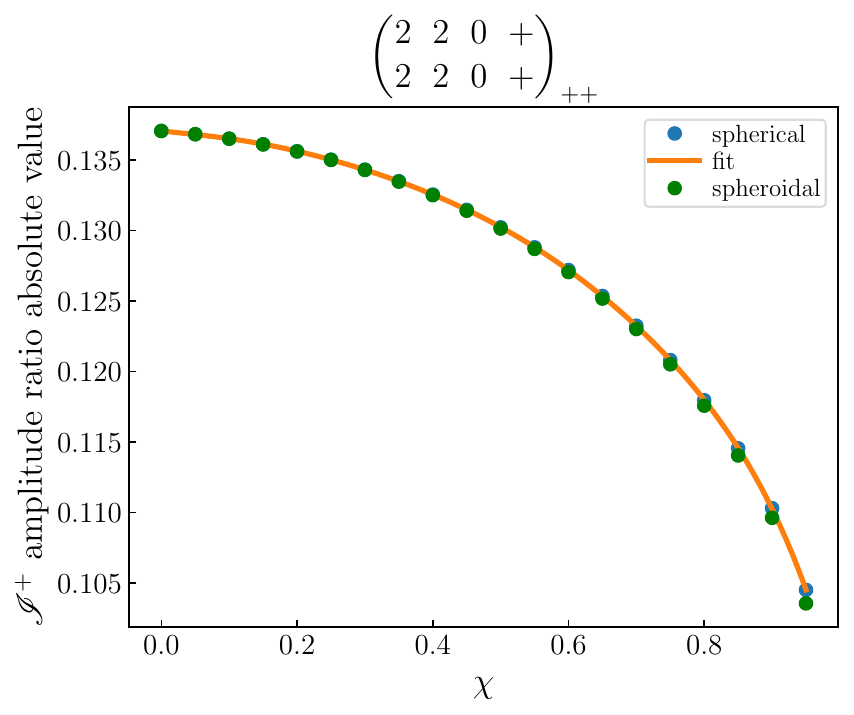
    }
    \includegraphics[width=\linewidth]{
        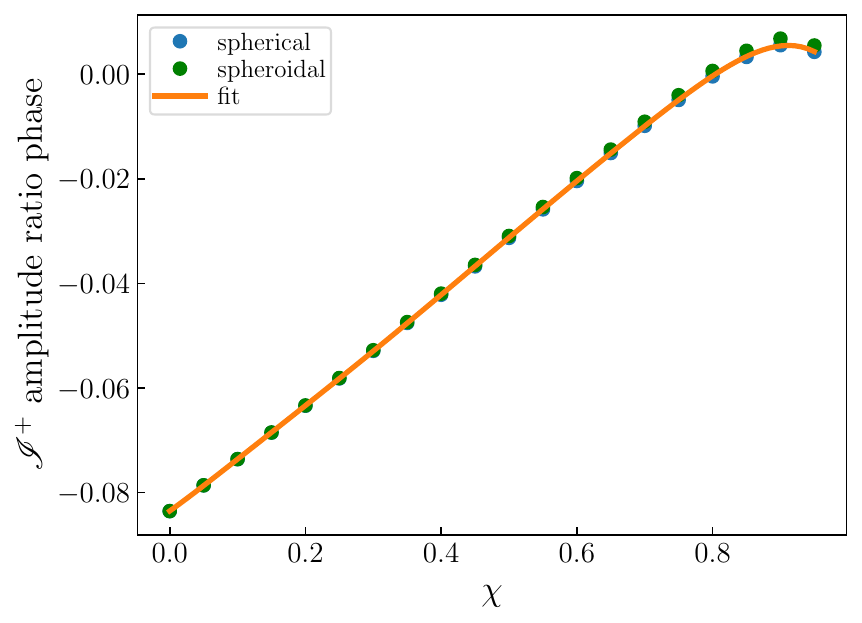
    }
    \caption{Magnitude (top) and phase (bottom) of $R_{++}$ for the QQNM
    $\begin{pmatrix}
        220+ \\
        220+
    \end{pmatrix}$ at null infinity, using spin-weighted spherical (blue dots) and
    spheroidal (green dots) harmonics}
    \label{fig:supplementary_spherical_vs_spheroidal_2201_2201}
\end{figure}

The value of $R$ depends on the choice of angular basis, whether using spin-weighted
spherical or spheroidal harmonics. In Fig.~\ref{fig:supplementary_spherical_vs_spheroidal_2201_2201},
we compare $R_{++}$ for $\begin{pmatrix}
    220+ \\
    220+
\end{pmatrix}$ between the two bases. We find the results differ by $\le 1\%$
even for dimensionless spins as large as $0.95$. Nonetheless, there are modes
where the difference is significant.

The QQNM $\begin{pmatrix}
    220+ \\
    200-
\end{pmatrix}$ needs more attention. 
Our code encounters issues for this as $\chi\to 0$, and the value of $R_{++}$ diverges.
We suspect that this is because in this limit the frequency approaches a purely decaying quantity, and thus it coincides with the branch cut along the imaginary axis. In the finite dimensional truncation the branch cut is replaced by several poles. Because at a pole the Teukolsky operator is non-invertible, near a pole the operator becomes ill-conditioned and numerically challenging to invert. This feature indicates that such quadratic effects require special treatment, which
we leave for future work.

\begin{figure}[!h]
    \includegraphics[width=\linewidth]{
        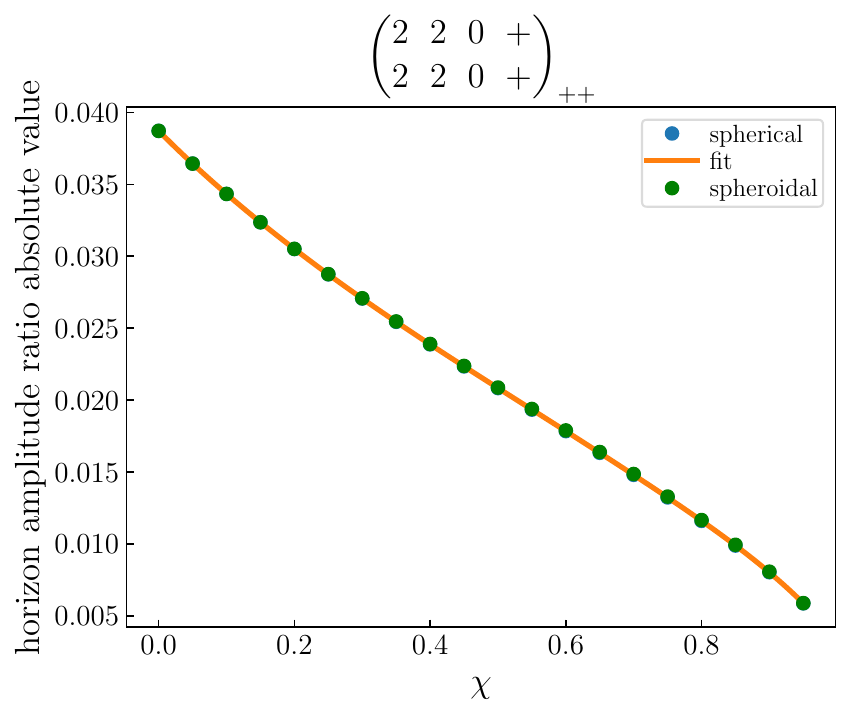
    }
    \includegraphics[width=\linewidth]{
        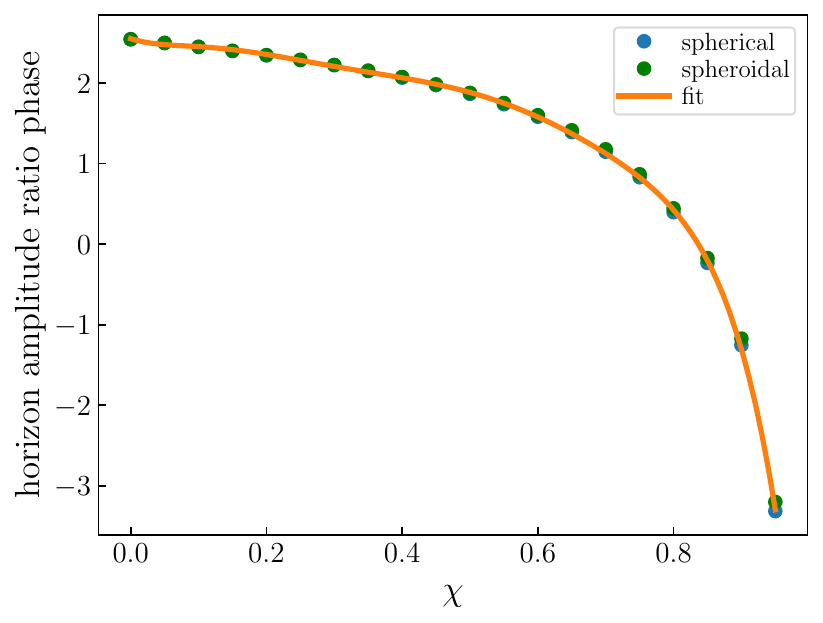
    }
    \caption{Magnitude (top) and phase (bottom) of the amplitude ratio at the
    horizon for the QQNM $\begin{pmatrix}
        220+ \\
        220+
    \end{pmatrix}$, using spin-weighted spherical (blue dots) and spheroidal (green
    dots) harmonics}
    \label{fig:horizon_2201_2201}
\end{figure}

Furthermore, the numerical code can also compute the amplitude of the nonlinear modes
at the horizon by solving the second order equation for $\Psi_{0}$. We
illustrate this in Fig.~\ref{fig:horizon_2201_2201} by plotting the amplitude of
the second order $\Psi_{0}$ for the $\begin{pmatrix}
    220+ \\
    220+
\end{pmatrix}$ QNM, divided by the amplitudes of the linear modes at null
infinity for normalization.

\subsection{Parity of linear and quadratic quasinormal modes}

In this section we will discuss the relationship between the positive and negative
frequency modes used in QNM analysis to even-parity (polar) and odd-parity (axial)
modes. First, we note that the parity transformation $P_{-}: (\theta, \phi) \to (
\pi - \theta, \phi + \pi)$ is a symmetry of the background Kerr spacetime.
Furthermore, the QNM boundary conditions are also preserved by the parity operator,
therefore QNMs are sent to QNMs.

Consider a linear mode $L = (\ell,\,m,\,n,\,p)$ with amplitude
$\mathcal{A}^{p}_{\ell m n}$. The parity transformation $P_{-}$ sends the waveform
$h(\theta, \phi)$ to $\bar{h}(\pi - \theta, \phi+\pi)$ \cite{Boyle:2014ioa}, the
waveform transforms as
\begin{align}
     & \mathcal{A}^{p}_{\ell m n}e^{-i \omega^p_{\ell m n} u}{}_{-2}S_{\ell m}(\theta, \phi; a \omega^{p}_{\ell m n})                                         \\
     & \to \bar{\mathcal{A}}^{p}_{\ell m n}e^{+i\bar{\omega}^p_{\ell m n} u}{}_{-2}\bar{S}_{\ell m}(\pi - \theta, \phi+\pi; a \omega^{p}_{\ell m n})\nonumber \\
     & = (-1)^{\ell+m}\bar{\mathcal{A}}^{p}_{\ell m n}e^{-i\omega^{-p}_{\ell, -m ,n} u}{}_{-2}S_{\ell(-m)}(\theta, \phi; a \omega^{-p}_{\ell (-m) n})\,,
\end{align}
where we have used Eq.~(1) from the Letter to rewrite the frequency, and
the following properties of spin weighted spheroidal harmonics~\cite{Vickers:2022ivk}:
\begin{align}
    {}_{s}\bar{S}_{\ell m}(\theta, \phi; a\omega)    & = (-1)^{s+m}{}_{-s}S_{\ell(-m)}(\theta, \phi; -a\bar{\omega}), \\
    {}_{s}S_{\ell m}(\pi -\theta, \phi+\pi; a\omega) & = (-1)^{\ell}{}_{-s}S_{\ell m}(\theta, \phi; a\omega)\,.
\end{align}
Therefore, the parity transformation takes the $L=(\ell, m, n, p)$ mode with amplitude
$\mathcal{A}^{p}_{\ell m n}$ to the mode $\bar{L}= (\ell, -m, n, -p)$ mode with amplitude
$(-1)^{\ell+m}\bar{\mathcal{A}}^{p}_{\ell m n}$. We will refer to $\bar{L}$ as
the conjugate mode of $L$. Furthermore, by taking a linear combination of the
linear modes $L$ and its conjugate mode $\bar{L}$, we can construct solutions that
are eigenvectors of the parity operator. In particular, the even parity sector is
when the eigenvalue is $(-1)^{\ell+m}$, and the odd parity sector is when the
eigenvalue is $(-1)^{\ell+m+1}$. The sign comes because the even and odd parity sectors
are traditionally separated using the sign arising from the reversal of orientation
rather than the parity operator $P_{-}$, giving rise to these signs.

Consequently, if we have the combination of $L=(\ell, m, n, p)$ and $\bar{L}=(\ell
, -m, n, -p)$ modes with amplitudes $\mathcal{A}_{\ell m n}^{p}=\bar{\mathcal{A}}
_{\ell(-m)n}^{-p}$, then combination of modes is even. If, on the other hand, $\mathcal{A}
_{\ell m n}^{p}= -\bar{\mathcal{A}}_{\ell(-m)n}^{-p}$, the combination of modes
is odd.

Consequently, an even or odd combination of pairs of linear modes $L_{1}$,
$\bar{L}_{1}$, and $L_{2}$, $\bar{L}_{2}$ gives rise to 4 quadratic modes that
arise from their cross-terms: $(L_{1}; L_{2})$, $(L_{1}, \bar{L}_{2})$, $(\bar{L}
_{1}; L_{2})$ and $(\bar{L}_{1}; \bar{L}_{2})$. Here each quadratic mode is
contributed to by 3 channels. However, as we will see, not all 12 of the amplitude
ratios are independent. The parity symmetry of the Kerr background will reduce
it to 6 independent quadratic ratios.

{\it Parity symmetry for quadratic amplitude ratios.---} Let us now use the
parity symmetry to relate amplitude ratios. Consider $Q =
\begin{pmatrix}
    \ell_{1} & m_{1} & n_{1} & p_{1} \\
    \ell_{2} & m_{2} & n_{2} & p_{2}
\end{pmatrix}_{++}$, a quadratic mode with frequency $\omega_{Q}$ produced using
the channel $(+, +)$. While the quadratic modes do not have spheroidal harmonic angular
dependence anymore, the strain at infinity can be decomposed into spheroidal
harmonics, so that the quadratic strain $h$ from this mode is given by
\begin{equation}
    \label{eq:quadratic-1-2}h_{Q}= \sum_{\ell^\prime}\mathcal{A}_{1}\mathcal{A}_{2}
    R^{Q,\ell^\prime}_{++}\;{}_{-2}S_{\ell^\prime (m_1+m_2)}(\theta, \phi; a\omega
    _{Q}) e^{-i\omega_Q u}\,.
\end{equation}
Where $R^{Q,\ell^\prime}_{++}$ is the amplitude ratio for the quadratic mode Q with
child harmonics $(\ell^{\prime}, m_{1}+ m_{2})$ with the $(+, +)$ channel,
$\mathcal{A}_{1}$ is the amplitude of the mode $(\ell_{1}, m_{1}, n_{1}, p_{1})$
and $\mathcal{A}_{2}$ is the amplitude of the mode
$(\ell_{2}, m_{2}, n_{2}, p_{2})$. Under a parity transformation, the linear
modes transform to $(\ell_{1}, -m_{1}, n_{1}, -p_{1})$ and
$(\ell_{2}, -m_{2}, n_{2}, -p_{2})$ with amplitudes
$(-1)^{\ell_1+m_1}\bar{A}_{1}$ and $(-1)^{\ell_2+m_2}\bar{A}_{2}$ respectively. Furthermore,
under the parity transformation, the quadratic mode~\eqref{eq:quadratic-1-2}
transforms to the quadratic mode $\bar{Q}=
\begin{pmatrix}
    \ell_{1} & -m_{1} & n_{1} & -p_{1} \\
    \ell_{2} & -m_{2} & n_{2} & -p_{2}
\end{pmatrix}_{++}$. We will refer to $\bar{Q}$ as the conjugate mode of $Q$.
\begin{align}
    h_{\bar{Q}}= \sum_{\ell^\prime} & (-1)^{\ell^\prime + m_1+m_2}\nonumber                                                                                                                                         \\
                                    & \times \bar{R}^{Q,\ell^\prime}_{++}\bar{\mathcal{A}}_{1}\bar{\mathcal{A}}_{2}e^{-i \omega_{\bar{Q}} u}{}_{-2}S_{\ell^\prime, -(m_1+m_2)}(\theta, \phi; a\omega_{\bar{Q}})\, .
\end{align}
Dividing the amplitude of this quadratic mode by the amplitudes of the linear
modes gives us the coupling coefficient $R^{\bar{Q},\ell^\prime}_{++}$ for the
$\bar{Q}$ mode, namely that
\begin{equation}
    \label{eq:mirror-coupling-coefficient}R^{\bar{Q}, \ell^\prime}_{++}= (-1)^{\ell^\prime
    + \ell_1 + \ell_2}\bar{R}^{Q, \ell^\prime}_{++}\,.
\end{equation}
Similarly, the above analysis applies to any channel $(c_{1}, c_{2})$, giving us
\begin{equation}
    R^{\bar{Q}, \ell^\prime}_{c_1 c_2}= (-1)^{\ell^\prime + \ell_1 + \ell_2}\bar{R}
    ^{Q, \ell^\prime}_{c_1 c_2}\,.
\end{equation}
Furthermore, note that while we used the spheroidal basis for this analysis, Eq.~\eqref{eq:mirror-coupling-coefficient}
is valid even for the amplitude ratio computed using the spin-weighted spherical
harmonic basis. Thus, the amplitude ratio of a quadratic mode can be expressed in
terms of the amplitude ratio of its conjugate mode.

\subsection{Additional SNR plots}

\begin{figure}
    \centering
    \includegraphics[width=\linewidth]{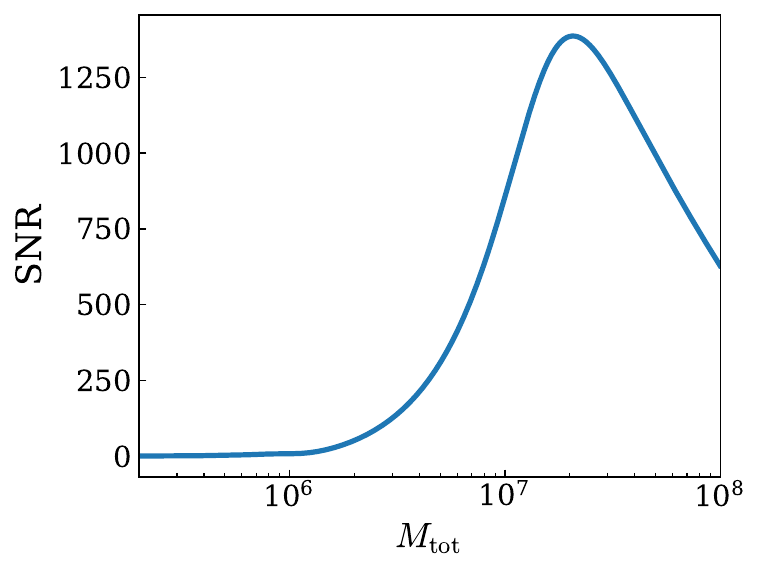}
    \includegraphics[width=\linewidth]{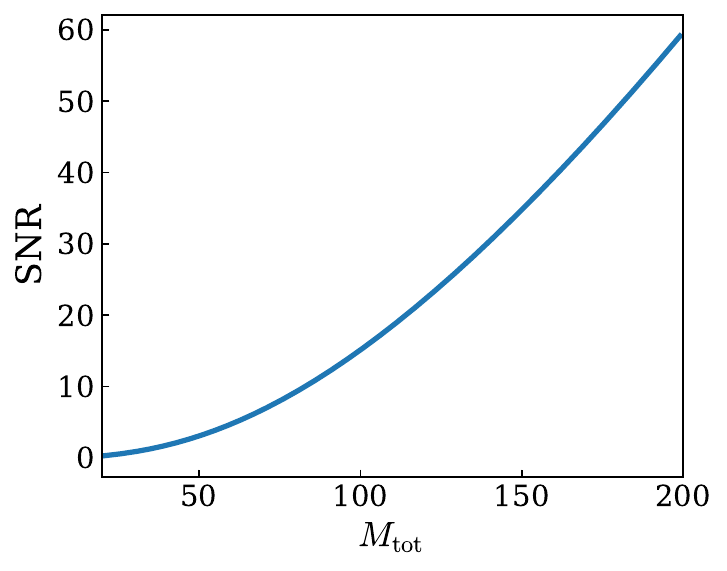}
    \caption{The SNR dependence of $\begin{pmatrix}
        3, & 3,  & 0, & + \\
        2, & 2, & 0, & +
    \end{pmatrix}$ on progenitor binaries' total mass $M_{\rm tot}$, using LISA
    (top) and CE (bottom), with $q=2.75,\chi_{\rm eff}=0.9$. For CE, we choose $D
    _{L}=410$ Mpc, while for LISA we set $z=1$.}
    \label{fig:SNR_3301_2201_mass}
\end{figure}

Figure \ref{fig:SNR_3301_2201_mass}
shows the SNR dependence of $\begin{pmatrix}
    3, & 3,  & 0, & + \\
    2, & 2, & 0, & +
\end{pmatrix}$ on progenitor binaries' total mass $M_{\rm tot}$, with
$q=2.75,\chi_{\rm eff}=0.9$. For CE, we choose $D_{L}=410$ Mpc, while for LISA we
set $z=1$.


\begin{figure}[tb]
        \centering
       {\includegraphics[width=\linewidth,clip=true]{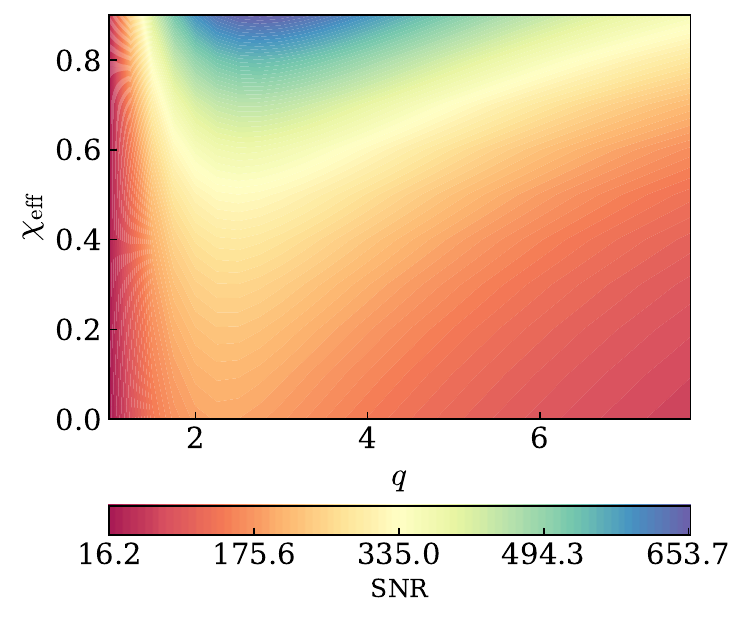}}
        \caption{The SNR dependence of the QQNM, $\begin{pmatrix}
            3, & 3, & 0, & + \\
            2, & 2, & 0, & +
        \end{pmatrix}$, on progenitor binaries' mass ratio $q$ and effective
        spin $\chi_{\rm eff}$, assuming LISA. The binary systems have a total mass of $8\times 10^{6}M_{\odot}$ and a redshift of 1. }
        \label{fig:330+_220+_SNR_2D_LISA}
    \end{figure}

    Figure \ref{fig:330+_220+_SNR_2D_LISA} shows the SNR dependence of $\begin{pmatrix}
    3, & 3,  & 0, & + \\
    2, & 2, & 0, & +
    \end{pmatrix}$ on $q$ and $\chi_{\rm eff}$, using LISA with $M_{\rm tot}=8\times 10^6M_\odot$ and a redshift of 1. The optimal binary configuration remains $q\sim 2.75$ and maximally positive $\chi_{\rm eff}$, as in CE.

\end{document}